\documentclass{PoS}

\usepackage{amsmath,bm,amssymb,color}
\usepackage{graphicx}
\usepackage{subcaption}
\usepackage{xspace}

\newcommand{\NNLOJET}{NNLO\protect\scalebox{0.8}{JET}\xspace}
\def\d{\hbox{d}}

\title{NNLO QCD Corrections for Higgs-plus-jet Production in the Four-lepton Decay Mode}

\ShortTitle{NNLO QCD Corrections for Higgs-plus-jet Production in the Four-lepton Decay Mode}

\author{\speaker{Xuan Chen}\\
University of Z\"urich\\
E-mail: \email{xuan.chen@uzh.ch}}

\author{Thomas Gehrmann\\
University of Z\"urich\\
E-mail: \email{thomas.gehrmann@uzh.ch}}

\author{Nigel Glover\\
University of Durham\\
E-mail: \email{e.w.n.glover@durham.ac.uk}}

\author{Alexander Huss\\
CERN\\
E-mail: \email{alexander.huss@cern.ch}}

\abstract{We present the computation of Higgs boson production in association with a jet at the LHC including QCD corrections up to NNLO. The calculation includes the subsequent decay of the Higgs boson into four leptons, allowing for the full reconstruction of the final-state kinematics. In anticipation of improved LHC measurements based on the full Run II dataset, we present a study for single- and double-differential cross sections within the fiducial volume as defined in prior ATLAS analyses. Higher-order corrections are found to have a sizeable impact on both normalisation and shape of differential cross sections.}

\FullConference{14th International Symposium on Radiative Corrections (RADCOR2019)\\ 
9-13 September 2019\\
		Palais des Papes, Avignon, France}

\begin{document}

\section{Introduction}
After the discovery of the Higgs boson \cite{Aad:2012tfa}, precise measurements of its properties become a focus of the LHC experiments. These measurements stress-test the Standard Model predictions for Higgs boson production for various decay channels as well as details of the underlying electroweak symmetry breaking mechanism. With the accumulation of data collected by LHC experiments, the measured properties of the Higgs boson in terms of mass, couplings and kinematic behaviours, are becoming increasingly precise. To analyse the state-of-the-art experimental results, equally precise theory predictions are mandatory and ideally adapted to the same final-state definition (fiducial selection criteria of the Higgs boson decay products and associated objects such as hadronic jets) used in the experimental analysis. The theoretical tools for these predictions are based on perturbation theory and implemented in Monte Carlo event generators, allowing to adapt to experimental fiducial cuts and becoming 
increasingly precise as higher order corrections are considered. 

At the LHC, the dominant Higgs boson production channel predicted by the Standard Model is gluon fusion \cite{Georgi:1977gs}. The Higgs boson is produced via quark-loop induced process from two parent gluons inside the colliding protons. In perturbative QCD calculations, the Born process (LO) is proportional to $\mathcal{O}(\alpha_s^2)$, yielding a Higgs boson with vanishing transverse momentum. Starting from next-to-leading order (NLO), the Higgs boson can be produced with finite transverse momentum by recoiling against QCD radiation. Large QCD corrections are observed at NLO \cite{Spira:1995rr} and an extensive study on the convergence of higher order corrections is motivated for the gluon fusion channel. To reduce the complexity of multi-loop calculations, the quark-loop-induced coupling between the Higgs field and gluons can be approximated in the limit of infinite top quark mass described by an effective field theory vertex \cite{Wilczek:1977zn}. In this limit, up to the third order of perturbation expansions (N3LO) at $\mathcal{O}(\alpha_s^5)$ is known for inclusive cross sections \cite{Dawson:1990zj, Anastasiou:2002yz, Catani:2007vq, Anastasiou:2015ema} and rapidity distributions \cite{Cieri:2018oms}. 

Precise measurements of Higgs properties typically require the identification of the production process. These types of analysis are obtained by constraining the associated objects produced with the Higgs boson. The underlying theory predictions are closely related to Higgs-plus-jet production, which starts only at $\mathcal{O}(\alpha_s^3)$ with the Higgs boson produced at finite transverse momentum. The first order in the perturbative expansion (NLO, $\mathcal{O}(\alpha_s^4)$) is known retaining the exact top quark mass dependence \cite{Lindert:2018iug, Jones:2018hbb,Neumann:2018bsx} while the NNLO corrections are only known within the heavy top mass limit \cite{Boughezal:2015dra, Boughezal:2015aha, Caola:2015wna, Chen:2016zka, Chen:2019wxf, Campbell:2019gmd}. Further efforts have been devoted to study the $H\rightarrow \gamma\gamma$ \cite{Caola:2015wna, Chen:2016zka}, $H\rightarrow 2l+2\nu$ \cite{Caola:2015wna} and $H\rightarrow 4l$ \cite{Chen:2019wxf} decay modes with selection algorithms used by ATLAS or CMS experiments for precise comparison and better understanding of acceptance factors. 

In this talk, we continue the precision study of  Higgs boson production  for 
the four-lepton decay mode $H\rightarrow ZZ^*\rightarrow 4l$. The light charged leptons provide a clean final state signature for the reconstruction of a Higgs boson at high transverse momentum. We apply the framework in \cite{Chen:2019wxf} including up to NNLO QCD corrections for the Higgs-plus-jet production (within the heavy top mass limit) and with Higgs-to-four-lepton decays. Using the kinematic information of the companying hadronic jet as a handle, we study kinematic properties of the Higgs boson through single and double differential cross sections. We use the parton-level event generator \NNLOJET \cite{antsub} with the fiducial cuts defined by the ATLAS analysis in \cite{ATLAS:2018bsg}. Our predictions can be directly compared with upcoming ATLAS measurements including the full 2016--18 dataset \cite{ATLAS:2019ssu}. 

\section{Implementation}
\label{Implementation}
At the LHC, the Higgs boson is mainly produced through gluon fusion induced by a quark-loop. The coupling strength of the quark loop is dominated by the top quark which can be integrated out, provided that the energy scales involved in the process are smaller than twice the top quark mass. In the heavy top mass limit (HTL), the Standard Model (SM) Lagrangian can be described by an effective Lagrangian. The matching between HTL and SM has been studied up to the four-loop order \cite{Chetyrkin:1997un,Kniehl:2006bg}. Within the HTL framework, fixed-order perturbative corrections can be calculated in massless QCD with five flavours. 

We study in the HTL framework the NNLO QCD corrections to Higgs boson production associated with a hadronic jet. We include all partonic channels contributing at NNLO, which are double-virtual (two-loop matrix elements of Higgs-plus-three-parton processes \cite{Gehrmann:2011aa}), real-virtual (one-loop matrix elements of Higgs-plus-four-parton processes \cite{Dixon:2009uk}) and double-real (tree-level matrix elements of Higgs-plus-five-parton processes \cite{DelDuca:2004wt}) contributions. Each of the three types of NNLO contributions is individually infrared divergent with respect to their phase space integration while the sum of all NNLO sub-processes is finite. We employ the antenna subtraction method to remove the infrared divergences for each sub-processes \cite{antsub}. For the Higgs-to-four-lepton decay mode, we apply the narrow-width approximation 
to the Higgs boson, using LO matrix elements for light charged-lepton final states ($4e$, $4\mu$ and $2e+2\mu$) \cite{Rizzo:1980gz}. Our calculation is implemented within to the parton-level event generator \NNLOJET.

For the numerical evaluation, we always assume an on-sell Higgs boson produced with a mass of $m_H=125$ GeV and the decay width of $\Gamma_H=4.1\times 10^{-3}$ GeV. We apply the $G_{\mu}$-scheme and fix the vacuum exception value to $v=246.2$ GeV. The mass of the Z-boson is $m_Z=91.1876$ GeV with decay width $\Gamma_Z=2.5$ GeV and the top quark mass (only appearing in the NNLO Wilson coefficient) is 173.2 GeV. For the choice of parton distribution functions (PDFs), we use a single set of PDFs (PDF4LHC15$\_$nnlo$\_$mc) \cite{Ball:2014uwa} with $\alpha_s(m_Z)=0.118$ for the LO, NLO and NNLO predictions. To estimate the theoretical uncertainties from QCD factorisation ($\mu_F$) and renormalisation ($\mu_R$) scale choices, we calculate each event with the customary seven-point variation around a dynamical central value defined as, 
\begin{equation}
  \label{eq:scale}
\mu \equiv \mu_R = \mu_F = \frac{1}{2}\sqrt{m_H^2 + (p^{4l}_T)^2} = \frac{1}{2} E_T^H,
\end{equation}
where $p^{4l}_T$ is the transverse momentum of the Higgs boson. The scale uncertainties presented in section \ref{Results} as bands are the envelopes of the seven scale choices. 

Precise measurements of differential cross sections in the Higgs-to-four-lepton decay channel are reported by ATLAS and CMS using 2016 \cite{Aaboud:2017oem, Sirunyan:2017exp}, 2016--17 \cite{ATLAS:2018bsg} and 2016--18 \cite{CMS:2019chr} datasets. For a precise comparison to the upcoming analysis by ATLAS using the 2016--18 dataset \cite{ATLAS:2019ssu}, we apply  the same event selection cuts and algorithms: the 1st and 2nd leading leptons have transverse momentum larger than 20 and 15 GeV; the 3rd and 4th leading leptons have transverse momentum larger than 10 and 5 GeV; all leptons must have rapidity within $\pm 2.7$; the angular separation between any two leptons has to satisfy $\Delta R(l_i,l_j)>0.1$. In order to reduce contaminations of leptons from hadronic decays, the same-flavour-opposite-charge-lepton (SFOC) pair having invariant mass closest to $m_Z$ must have $m^{Z_1}_{l^+ l^-}\in [50, 106]$ GeV; the remaining SFOC pair must have $m^{Z_2}_{l^+ l^-}\in [12, 115]$ GeV. In the $4e$ or $4\mu$ final states, the additional two SFOC pairs need to satisfy $m_{l^+ l^-} > 5$ GeV. To identify final state jets, we use the anti-k$_T$ jet algorithm with radius parameter $R=0.4$. A jet can only be identified within the fiducial region of $p_T^j>30$ GeV and $|y^j|<4.4$. We also require each jet candidate to be well separated from leptons with $\Delta R(j,l)> 0.1$. 

With increasing statistics collected by ATLAS and CMS, a larger kinematic region is explored to study properties of the Higgs boson. For example, the measured Higgs transverse momentum distribution is extended above 350 GeV by the ATLAS analysis \cite{ATLAS:2019ssu}. The energy scales involved in describing Higgs properties in those energetic fiducial regions are comparable to or larger than twice of the top quark mass. This implies that the HTL approximation is no longer accurate for precision phenomenology and that top quark mass effects can not be neglected. The QCD corrections with exact top quark mass dependence for Higgs-plus-jet final states are calculated analytically at LO \cite{Ellis:1987xu} and numerically at NLO \cite{Jones:2018hbb}. Due to the complexity of extending the exact top quark mass dependence to NNLO, in the current study using \NNLOJET, we estimate the quark mass effects in the higher-order HTL predictions through re-weighting procedures. Through detailed comparisons of inclusive \cite{Spira:1995rr} and differential observables \cite{Jones:2018hbb}, NLO HTL re-weighted by full LO predictions provides a  well-improved approximation to the full NLO calculations. Different re-weighting procedures are compared in \cite{Chen:2016zka}, and as a default we apply a multiplicative rescaling to NLO and NNLO HTL by full LO predictions:
\begin{equation}
  \label{eq:Hto4lrewt}
 \frac{\d \sigma_{\text{N(N)LO}\otimes \text{LOM}}}{\d \mathcal O}= R_{\text{LO}}(\mathcal O)\bigg( \frac{\d \sigma^{\text{HTL}}_{\text{N(N)LO}}}{\d \mathcal O}\bigg),
\end{equation}
with
\begin{equation}
    R_{\text{LO}}(\mathcal O)= 
\begin{cases}
    \bigg(\frac{\d \sigma^{\text{M}}_{\text{LO}}}{\d \mathcal O}\bigg)\bigg/\bigg(\frac{\d \sigma^{\text{HTL}}_{\text{LO}}}{\d \mathcal O}\bigg),& \text{if } \frac{\d \sigma^{\text{HTL}}_{\text{LO}}}{\d \mathcal O}\ne 0\\
    1,              & \text{otherwise.}
\end{cases}
\end{equation}
We further include the bottom and charm quark mass effects in $\d \sigma^{\text{M}}_{\text{LO}}$ with $m_b=4.18$ GeV, $m_c=1.275$ GeV while keeping other light quarks massless.

\section{Results}
\label{Results}
In our previous paper \cite{Chen:2019wxf} we calculated Higgs and Higgs-plus-jet production in the Higgs-to-four-lepton decay mode based on studies by ATLAS \cite{Aaboud:2017oem, ATLAS:2018bsg} and CMS \cite{Sirunyan:2017exp, CMS:2019chr}. We investigated fiducial cross sections and several differential distributions with up to NNLO QCD corrections using the HTL approximation re-weighted by the LO including the full top mass dependence. The inclusion of NNLO corrections significantly improves the accuracy of the differential predictions with residual scale uncertainties below 20\% and a better describes the experimental data. With the experimental data currently having still large uncertainties \cite{CMS:2019chr} in the Higgs-to-four-lepton decay mode, the upcoming ATLAS analysis containing full Run II statistics \cite{ATLAS:2019ssu} is expected to provide more details on the properties of the Higgs boson.

Here we present further fiducial total and differential cross section predictions based on the ATLAS selection criteria \cite{ATLAS:2018bsg, ATLAS:2019ssu}.
Applying the fiducial cuts specified in section \ref{Implementation}, the total cross sections for Higgs-plus-jet produciton in Higgs-to-four-lepton decay channel are
\begin{equation}
\sigma^{4l+jet}_{LO}=0.621^{+0.256}_{-0.165}\ \text{fb},\qquad \sigma^{4l+jet}_{NLO}=1.04^{+0.196}_{-0.179}\ \text{fb},\qquad \sigma^{4l+jet}_{NNLO}=1.13^{+0.006}_{-0.073}\ \text{fb}.
\end{equation}
To include finite top mass dependence at LO, the re-weighted total cross sections are
\begin{equation}
\sigma^{4l+jet}_{LOM}=0.639^{+0.263}_{-0.170}\ \text{fb},\qquad \sigma^{4l+jet}_{NLO\otimes LOM}=1.07^{+0.203}_{-0.184}\ \text{fb},\qquad \sigma^{4l+jet}_{NNLO\otimes LOM}=1.16^{+0.006}_{-0.081}\ \text{fb}.
\end{equation}
The uncertainties of the above total cross sections are obtained from taking the envelope of the standard seven-scale variation. We observe a good convergence of fixed-order perturbative expansion going from NLO to NNLO. The finite top mass effect at LO amounts to an increase of about 3\% of  the HTL approximation. 

\begin{figure}
    \centering
    \begin{subfigure}[b]{0.49\textwidth}
        \includegraphics[width=\textwidth]{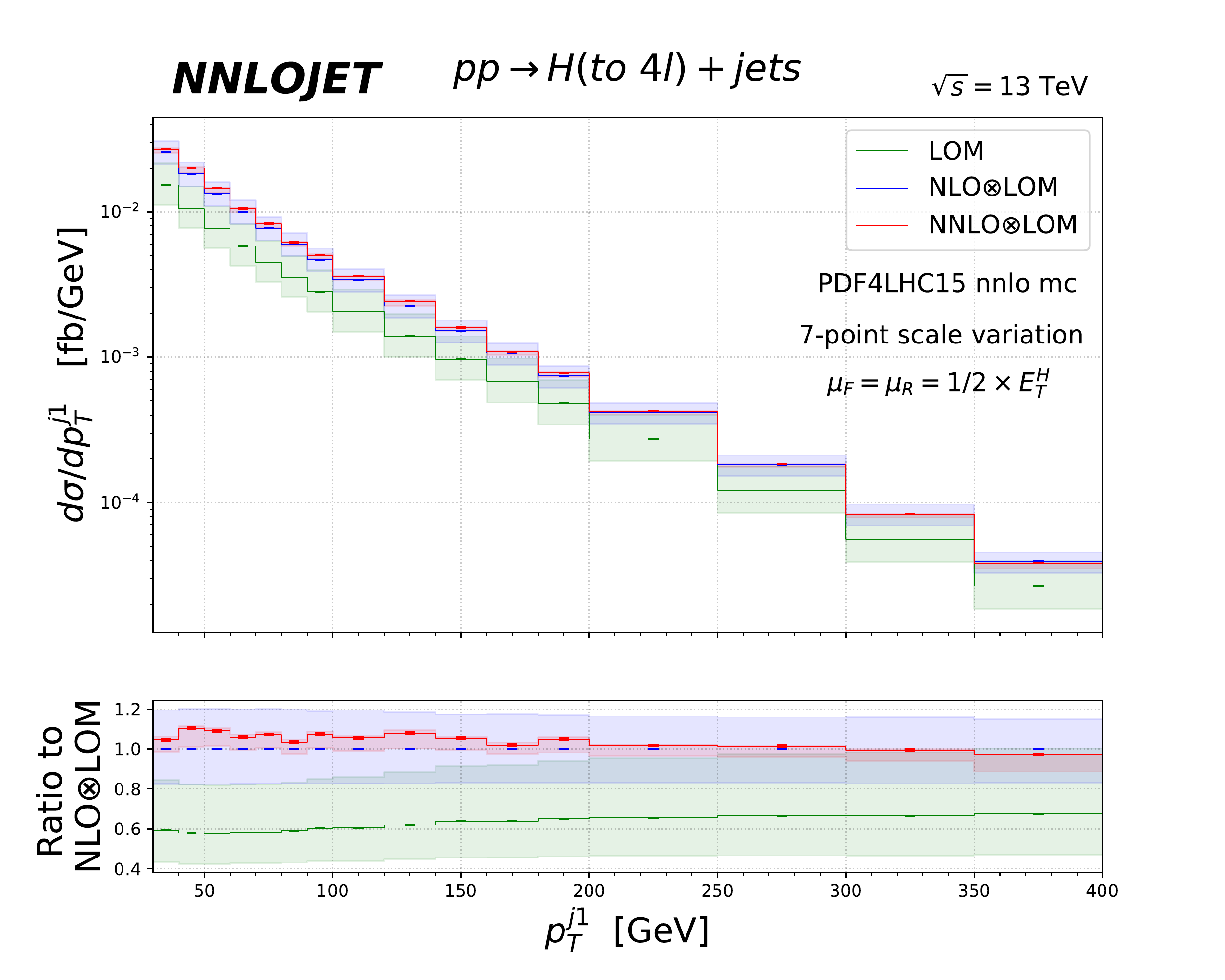}
    \end{subfigure}
    \begin{subfigure}[b]{0.49\textwidth}
        \includegraphics[width=\textwidth]{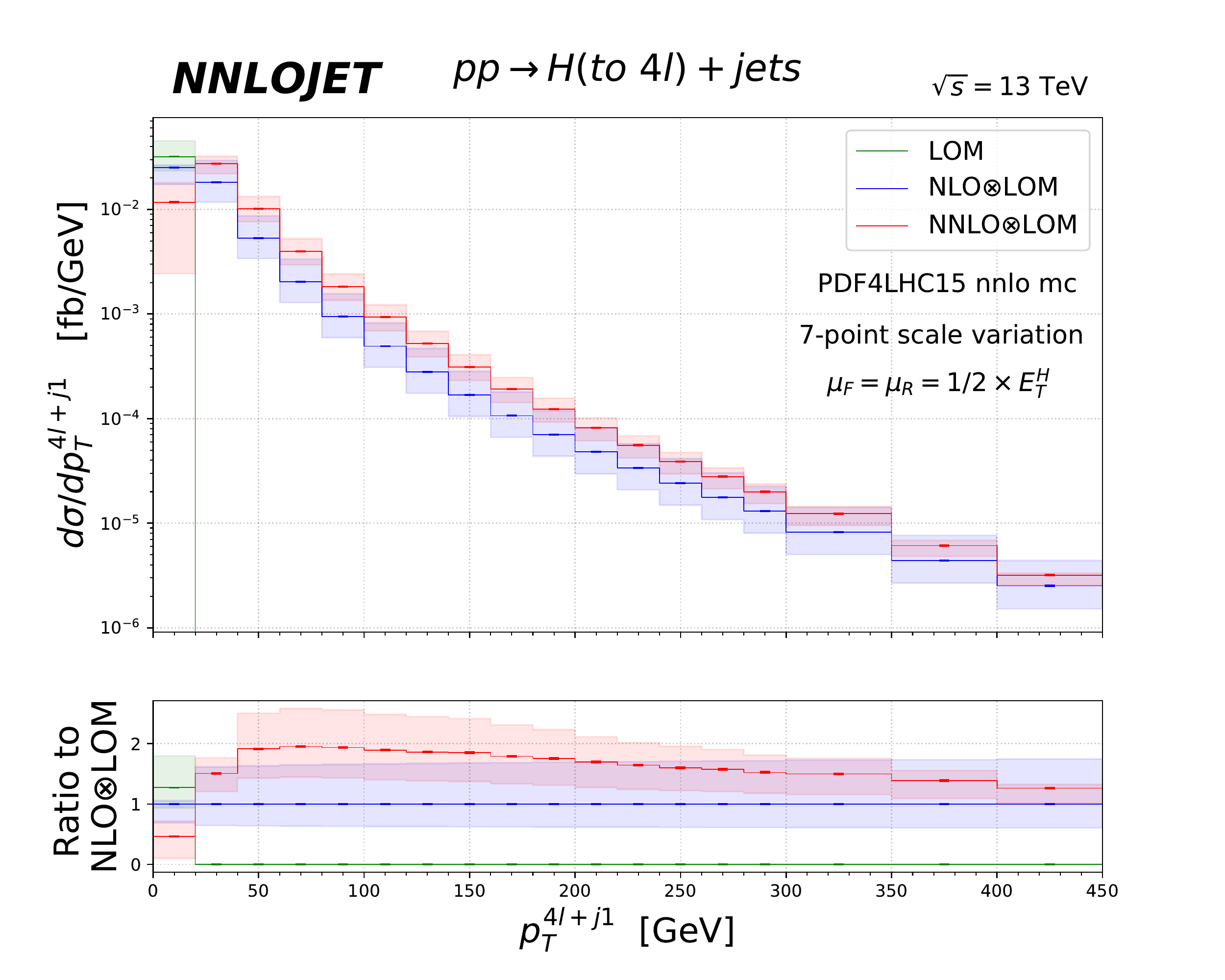}
    \end{subfigure}

    \begin{subfigure}[b]{0.49\textwidth}
        \includegraphics[width=\textwidth]{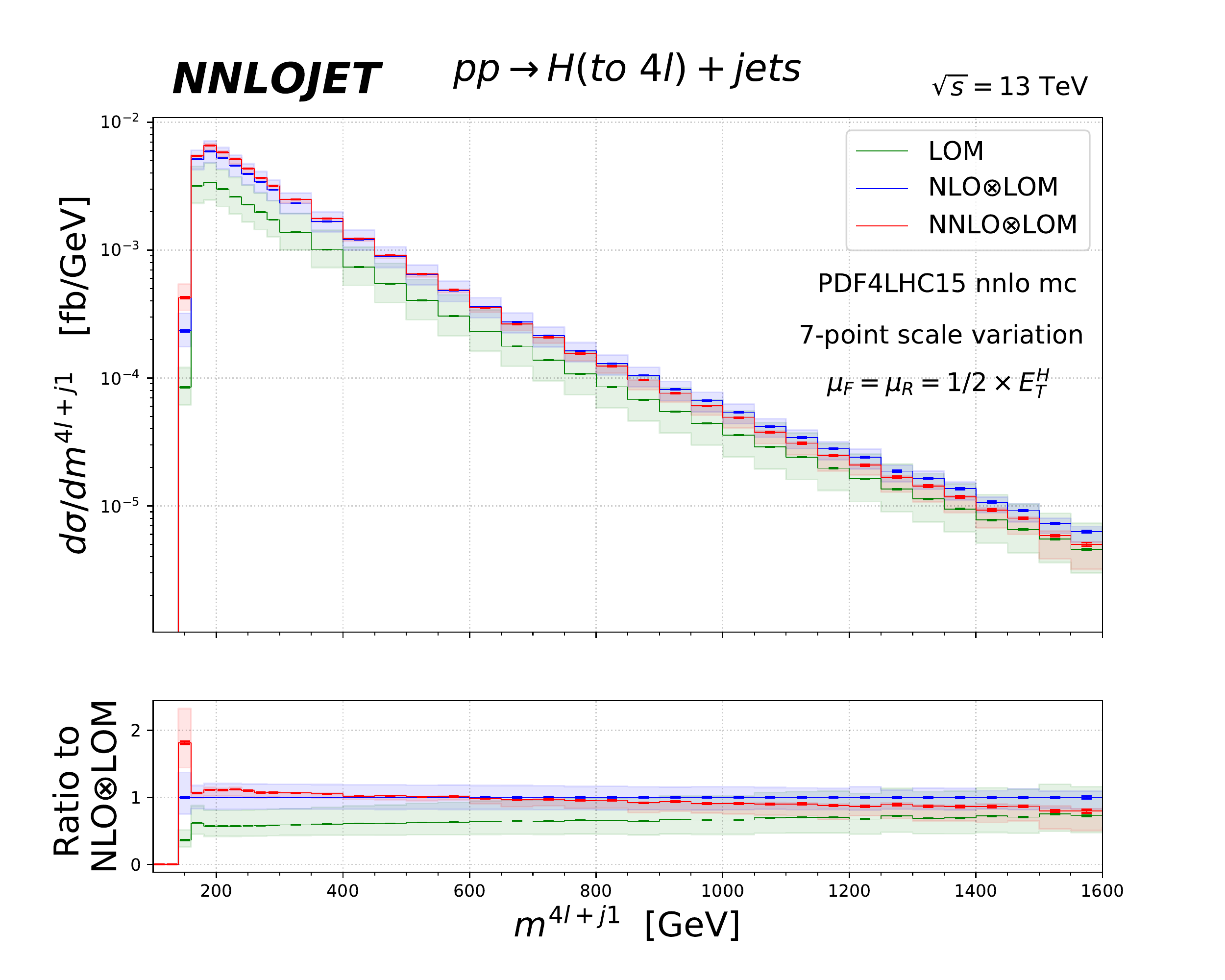}
    \end{subfigure}
    \begin{subfigure}[b]{0.49\textwidth}
        \includegraphics[width=\textwidth]{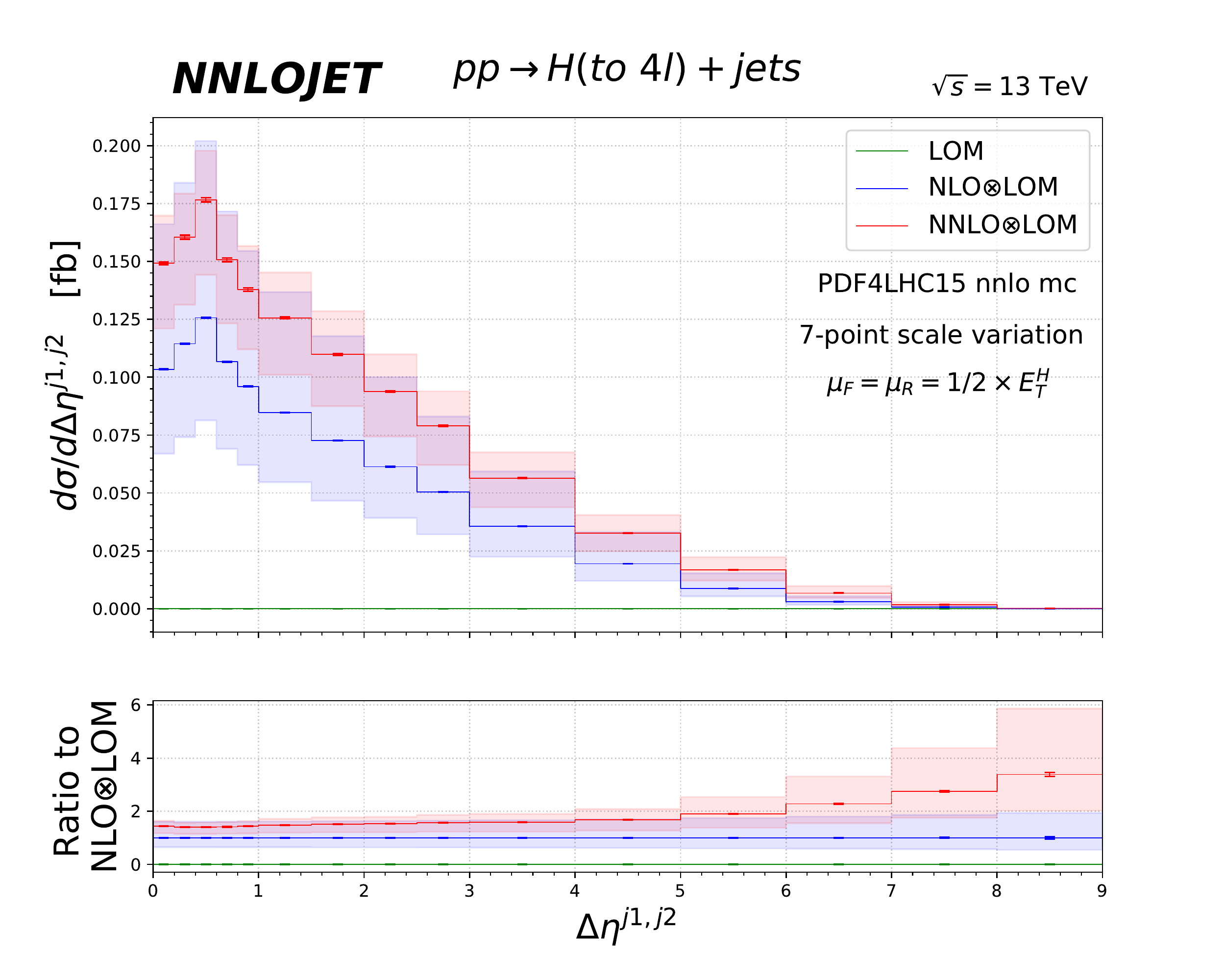}
    \end{subfigure}
    \caption{Fiducial single differential cross sections for the transverse momentum of the leading-jet (upper left), the transverse momentum of the system of four leptons plus the leading-jet (upper right), the invariant mass of the system of four leptons plus the leading-jet (lower left) and the pseudorapidity difference between the leading and the second-leading jet (lower right).}\label{fig:1D}
\end{figure}

In figure \ref{fig:1D}, we present results for fiducial single differential cross sections. For the transverse momentum distributions of the leading jet (upper left), we observe large NLO corrections of about $+80\%$ with respect to LO with a mild shift of the shape of the distribution. The NNLO corrections are consistent within the NLO scale variation band however result in a significant reduction of scale uncertainties across the entire considered kinematic range.
 
The invariant mass distributions of the four-lepton-plus-leading-jet system (lower left) contain no contribution below 140 GeV due to the jet identification requirement of $p^j_T > 30$ GeV. Close to the Higgs-plus-leading-jet threshold, we observe large QCD contributions of almost 100\% for each higher order with increasing scale variation bands. This indicates the breakdown of fixed order calculations near the production threshold and requires a more detailed study using threshold resummation. Away from the threshold, we observe large but flat NLO corrections with respect to LO while the NNLO correction modifies the shape of the distribution and reduces the scale uncertainties to about 10\% in the bulk of the distribution.

The transverse momentum distribution of the four-lepton-plus-leading-jet system (upper right) is a ``zero-bin'' observable constrained by transverse momentum conservation for Higgs-plus-jet production at LO. From NLO and above, the differential cross sections start to spread out to a wide dynamic range. Our results present large NNLO corrections to both the normalisation and the shape of the distributions with about $+100\%$ contributions with respect to NLO in the bulk of the cross sections. We also observe NNLO scale variation being larger than NLO below the jet production threshold of 30 GeV. The corresponding resummation effects are investigated in \cite{Sun:2014lna}. 

The distribution for the pseudorapidity difference between the leading and the second-leading jet is presented in the lower right part of figure \ref{fig:1D}. Events from LO Higgs-plus-jet production do not contribute to this observable and the highest nominal fixed-order corrections in our calculation are effectively only NLO-accurate. We observe that the peak of the distributions is centred around 0.5. For the bulk of the distribution with pseudorapidity difference below 2, the NNLO corrections stay inside the NLO scale variation band with scale uncertainties of about 30\% which is less than half of the uncertainties at NLO. Towards the tail of the distribution, our results indicate increasing NNLO contributions outside the NLO uncertainty band with up to 240\% corrections compared to NLO. The current results indicate the necessity to study Higgs-plus-two-jet production at NNLO to test the convergence of higher-order QCD corrections.

\begin{figure}
    \centering
    \begin{subfigure}[b]{0.49\textwidth}
        \includegraphics[width=\textwidth]{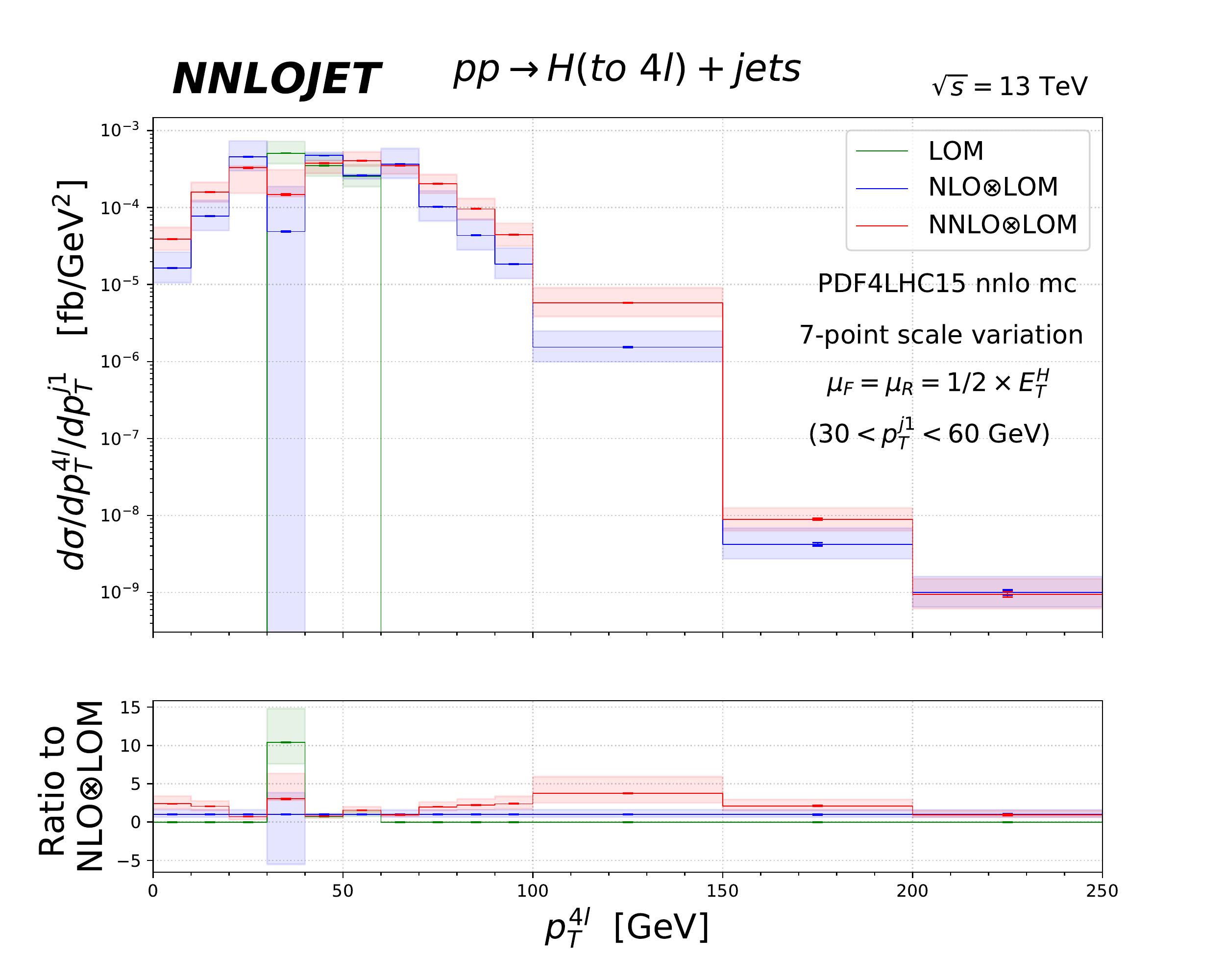}
    \end{subfigure}
    \begin{subfigure}[b]{0.49\textwidth}
        \includegraphics[width=\textwidth]{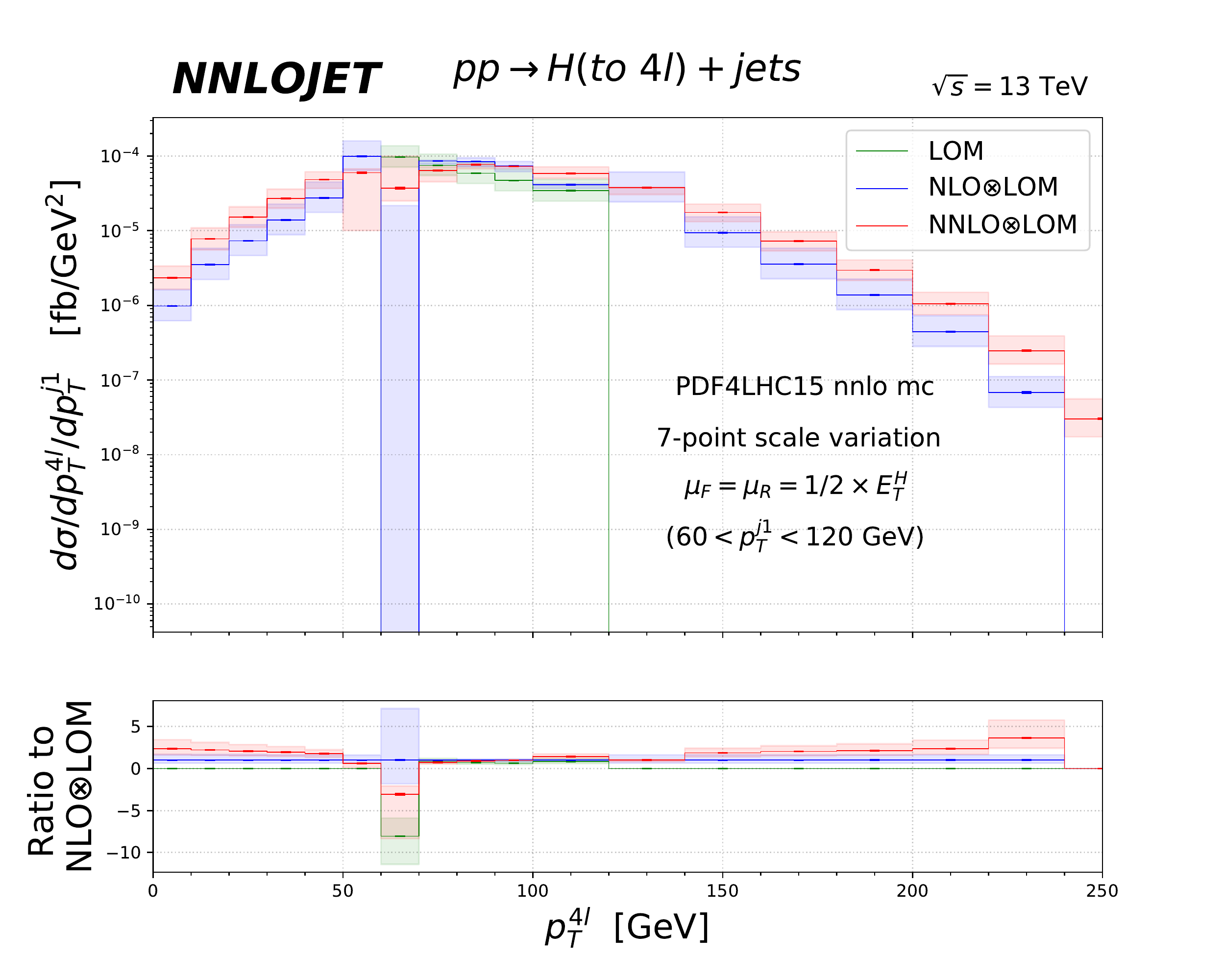}
    \end{subfigure}

    \begin{subfigure}[b]{0.49\textwidth}
        \includegraphics[width=\textwidth]{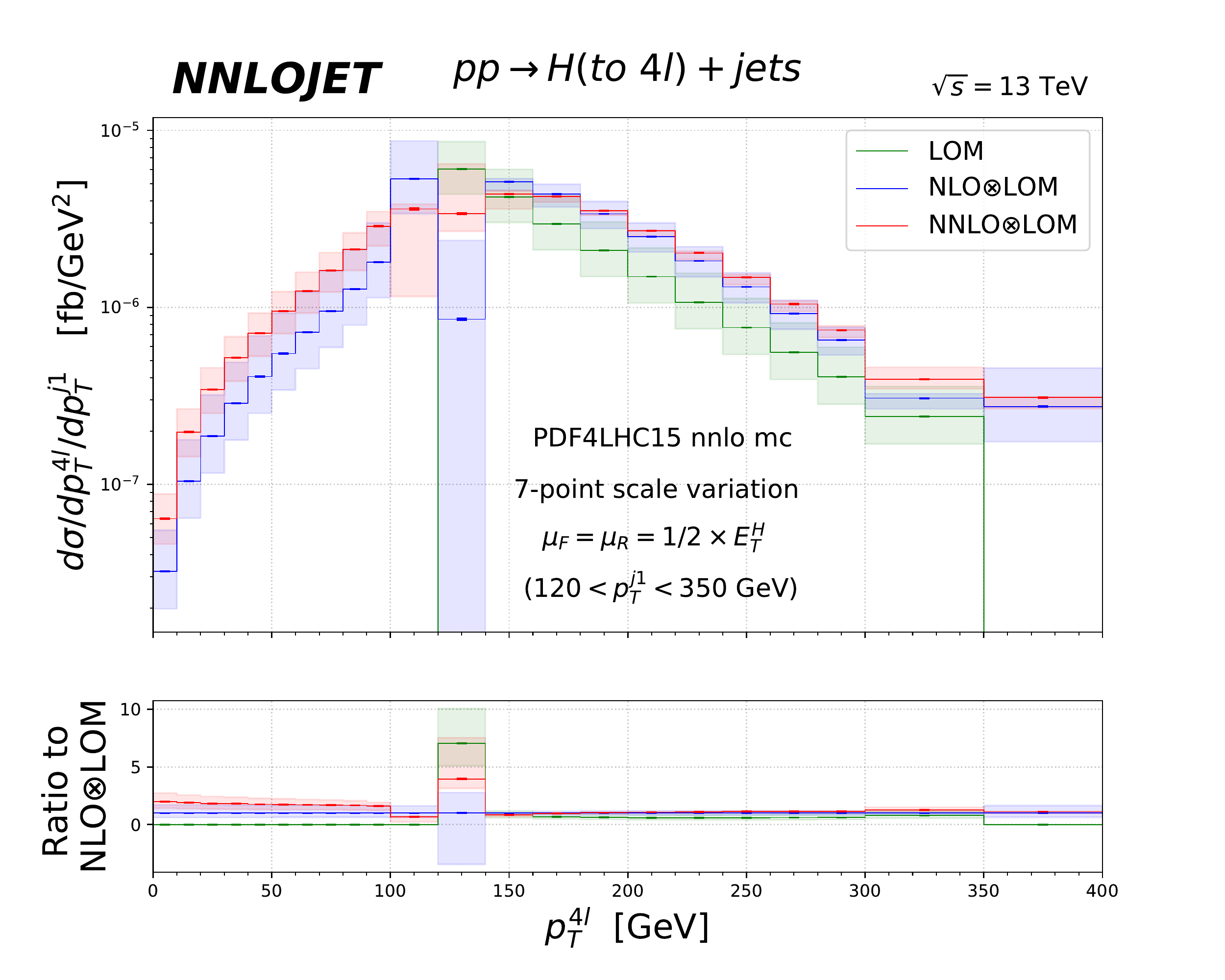}
    \end{subfigure}
    \caption{Fiducial double differential cross sections for the transverse momentum of the four-lepton system with respect to the transverse momentum of the leading jet.}\label{fig:2D}
\end{figure}

In figure \ref{fig:2D}, we present results for fiducial double differential cross sections for the transverse momentum of the four-lepton system ($p_T^{4l}$) in three dynamic regions for the transverse momentum of the leading jet ($p_T^{j1}$). We use the $p_T^{j1}$ observable as a handle to select events produced with energetic Higgs bosons. An event is discarded if it fails the constraint on $p_T^{j1}$. With transverse momentum conservation, the LO Higgs-plus-jet predictions only contribute inside the fiducial region of the $p_T^{j1}$ cut. For the histogram bins adjacent to the $p_T^{j1}$ cuts of 30, 60, 120 and 350 GeV, we observe the well-known Sudakov shoulder effects distorting the central scale predictions and the scale variation bands of fixed order calculations. We also notice that the distortion is much more severe at the lower end of the $p_T^{j1}$ cut compared to the higher end. More work on the resummation of double differential $p_T^{4l}$ distributions in Higgs-plus-jet final states is needed in the future. Away from the $p_T^{j1}$ cuts, the NNLO corrections are positive with increasing k-factors towards both ends of the distributions. Within the $p_T^{j1}$ cut region, we observe a nice convergence of scale variation bands when including higher order corrections. Outside the $p_T^{j1}$ cut region, our highest nominal fixed order corrections are effectively only  NLO-accurate and we do not yet observe evidence for the convergence of the perturbative series.

\section{Conclusions}
In this proceedings article, we studied the Higgs-plus-jet production in the $H\rightarrow 4l$ decay mode within the fiducial cuts defined by the ATLAS analysis \cite{ATLAS:2018bsg, ATLAS:2019ssu}. We include in the produciton part the QCD corrections up to NNLO and illustrate the effects from higher-order corrections in fiducial total and differential cross sections. We used the parton-level event generator \NNLOJET to preform the calculation in the heavy top mass limit. Finite top mass effects were approximately accounted for through a leading-order multiplicative re-weighting. 

We found sizable NNLO corrections to fiducial total cross sections and to various kinematic regions of single- and double-differential distributions. Including the second order of the perturbative expansion improves most predictions with a substantial reduction of scale uncertainties to a level of about 10\%. Despite the non-flat NNLO k-factors, convergence of perturbative expansions is indicated by the overlap of scale variation bands between NLO and NNLO. To study properties of energetic Higgs bosons, we presented single- and double-differential cross sections using kinematic information of the accompanying jets to identify the recoiling Higgs boson. In the kinematic regions where our highest nominal fixed order corrections only contribute to the first order of perturbative expansions, we demonstrated significant NNLO corrections outside the NLO scale uncertainty bands.

Our results prepare the benchmark predictions of detailed properties of the Standard Model Higgs boson. With a clean environment to reconstruct final state leptons, our results are ready to be compared with upcoming LHC analysis including the full Run II dataset \cite{ATLAS:2019ssu}.

\end{document}